\documentclass[prb,onecolumn,preprint,floatfix]{revtex4-1}

\usepackage {graphicx}
\usepackage [centertags]{amsmath}
\usepackage {amsfonts}
\usepackage {amssymb}
\usepackage {bm}
\usepackage{dcolumn}

\usepackage {epstopdf}

\newcommand{\ang}{~\text{\AA}}
\renewcommand{\vec}[1]{\boldsymbol{#1}}
\newcommand{\figwidth}{0.60\textwidth}

\begin{document}

\title{A Luttinger Liquid Core Inside Helium-4 Filled Nanopores}

\author{Adrian \surname{Del Maestro}}

\affiliation{Department of Physics, University of Vermont,
Burlington, VT 05405, USA}

\date{\today}

\begin{abstract}
As helium-4 is cooled below 2.17 K it undergoes a phase transition to a
fundamentally new state of matter known as a superfluid which supports flow
without viscosity.  This type of dissipationless transport can be observed by
forcing helium to travel through a narrow constriction that the normal liquid
could not penetrate. Recent experiments have highlighted the feasibility of
fabricating smooth pores with nanometer radii, that approach the truly one
dimensional limit where it is believed that a system of bosons (like helium-4)
may have startlingly different behavior than in three dimensions. The one
dimensional system is predicted to have a linear hydrodynamic description known
as Luttinger liquid theory, where no type of long range order can be sustained.
In the limit where the pore radius is small, Luttinger liquid theory would
predict that helium inside the channel behaves as a sort of quasi-supersolid
with all correlations decaying as power-law functions of distance at zero
temperature.  We have performed large scale quantum Monte Carlo simulations of
helium-4 inside nanopores of varying radii at low temperature with realistic
helium-helium and helium-pore interactions. The results indicate that helium
inside the nanopore forms concentric cylindrical shells surrounding a core that
can be fully described via Luttinger liquid theory and provides insights into
the exciting possibility of the experimental detection of this intriguing low
dimensional state of matter.  
\end{abstract}

\maketitle

\section{Introduction}

As the spatial dimension of an interacting many-body system is reduced, both
thermal and quantum fluctuations are enhanced, providing a fascinating arena for
the study of complex phenomena.  It is well known that in strictly one
dimension, there is no broken continuous symmetry but instead the persistence of
only quasi-long range order characterized by power-law decay of correlation
functions. The simplest example of a model displaying these features is the one
dimensional (1d) non-relativistic Galilean invariant Bose gas with delta
function interactions.  The ground state of this model has been known for almost
fifty years and can be solved exactly using the Bethe Ansatz\cite{Lieb:1963ik}.
An alternative approach, based on the concept of a low dimensional harmonic
fluid\cite{Tomonaga:1951kv,Mattis:1965dv,Luther:1974jl,Efetov:1975kb} was first
understood to be universal by Haldane\cite{Haldane:1981gd,Haldane:1981bk} with
the techniques formalized in these seminal works now generally known as
bosonization and Luttinger liquid (LL) theory.  The combination of these two
approaches indicate that there is no well defined quantum phase transition at
zero temperature as a function of the strength of the interactions in 1d.  As
interactions are increased, the system crosses over from a regime dominated by
phase fluctuations, to one with tendencies towards density wave order.
Likewise, there is no phase transition as a function of temperature. 

Historically, the experimental study of low dimensional physics has been
confined to fermionic systems such as insulating spin chains and quantum
wires\cite{Giamarchi:2004qp} and a bosonic realization has been lacking.  The
identification of accessible bosonic systems is particularly relevant in light
of the universal predictions of LL theory.  Recently, weakly interacting
ultra-cold bosonic gases consisting of laser trapped Alkali atoms confined to
cigar shaped quasi-one-dimensional condensates\cite{Greiner:2001cm} or
rings\cite{Ryu:2007fm} have been studied but it is difficult to obtain results
at higher densities where the constituent atoms are more strongly interacting.
The various theoretical approaches and considerations required to study such
systems have been covered in detail in a recent review
article\cite{Cazalilla:2011dm}. 

A considerably different approach involves the preparation of a quantum fluid of
interacting bosons, such as neutral ${}^4$He in a \emph{physically} confining
geometry at high density. In fact, the ability of helium-4 cooled below
$T_{\lambda} = 2.17$~K (the bulk superfluid transition temperature in ${}^4$He)
to flow through a narrow constriction (superleak) is one of the original
defining characteristics of the superfluid state of matter.  Original
experiments in this vein achieved physical confinement of helium through the
quasi-1D cavernous networks of porous glasses such as Vycor\cite{Beamish:1983co}
and more recently using folded sheets of mesoporous
materials\cite{Taniguchi:2010dh}.  Advances in nanofabrication techniques have
culminated in a more systematic approach that has been exploited by Savard
\emph{et al.}\cite{Savard:2009eq,Savard:2011fe} in studying $^4$He inside
nanopores of variable radii by sculpting a pore through a Si$_3$N$_4$ membrane
using a transmission electron beam.  Although these experiments have thus far
focused on flow properties of helium in the gas phase in
nanopores\cite{Savard:2009eq} and the superfluid phase for wide
nanoholes\cite{Savard:2011fe} they provide a tantalizing road map for the
experimental detection of a bosonic Luttinger liquid.  It is thus natural to
ponder a simpler equilibrium system, that can be studied numerically as a
function of pore radius and temperature below $T_{\lambda}$. We expect that
when the pore radius is sufficiently small, the length sufficiently long and the
temperature low enough with respect to $T_{\lambda}$ the ${}^4$He filled
nanopore should begin to exhibit strictly 1d behavior where LL theory can be
used to characterize the nature of the quasi-long-range order.  

The remainder of this paper is concerned with the precise definition of
\emph{sufficiently} in the previous sentence and is organized as follows.  We
first identify the particulars of a theoretical model of confined helium-4 and
introduce the stochastically exact numerical method employed to study its
behavior.  The numerical results are used to determine a phase diagram for
helium inside the pore with special attention payed to the types of structures
that are allowed by the Galilean invariant cylindrically symmetric confining
potential.  Density-density correlations inside the inner region of the pore are
then analyzed in terms of Luttinger liquid theory which is introduced in an
economical fashion with references provided to more complete and elaborate
treatments elsewhere.  The results highlight the applicability of the linear
hydrodynamics of LL theory in describing helium inside nanopores and a
discussion of the consequences and limitations of this finding is presented.

\section{Path Integral Simulations of Confined Helium-4}

The finite extent and translational invariance of the van der Walls interaction
between ${}^4$He atoms combined with their fundamental indistinguishably
conspire to make the numerical study of quantum fluids a challenging task.
Unbiased stochastically exact simulations of the quantum bosonic many-body
problem in the continuum at low temperature have only recently become more
feasible through the introduction of the continuous space Worm Algorithm
(WA)\cite{Boninsegni:2006ed,Boninsegni:2006gc}.  This method builds upon the
conventional Path Integral Monte Carlo approach of
Ceperley\cite{Ceperley:1995gr} that exploits the Feynman path integral
interpretation of quantum mechanics to perform Metropolis sampling of the
\emph{worldlines} of particles in $d+1$ dimensions.  The WA extends the
configuration space to include open worldlines known as \emph{worms} that are
not periodic in imaginary time and directly contribute to the single particle
Matsubara Green function allowing for simulations to be performed in the grand
canonical ensemble.  The simulation cell is kept in thermal equilibrium with a
surrounding bath with which it can exchange particles and the chemical potential
$\mu$ is an input parameter providing access to important physical observables
such as the compressibility.  In addition, the WA employs an efficient method
via the swap operator\cite{Boninsegni:2006ed} to directly sample the quantum
statistics of identical bosons through only local exchanges of worldlines.  Such
exchanges are the crucial step en route to the superfluid state of matter, with
the superfluid density being measured through a topological winding number;  the
standard technique in continuum simulations\cite{Pollock:1987ta}.

\begin{figure}
\includegraphics[width=\figwidth]{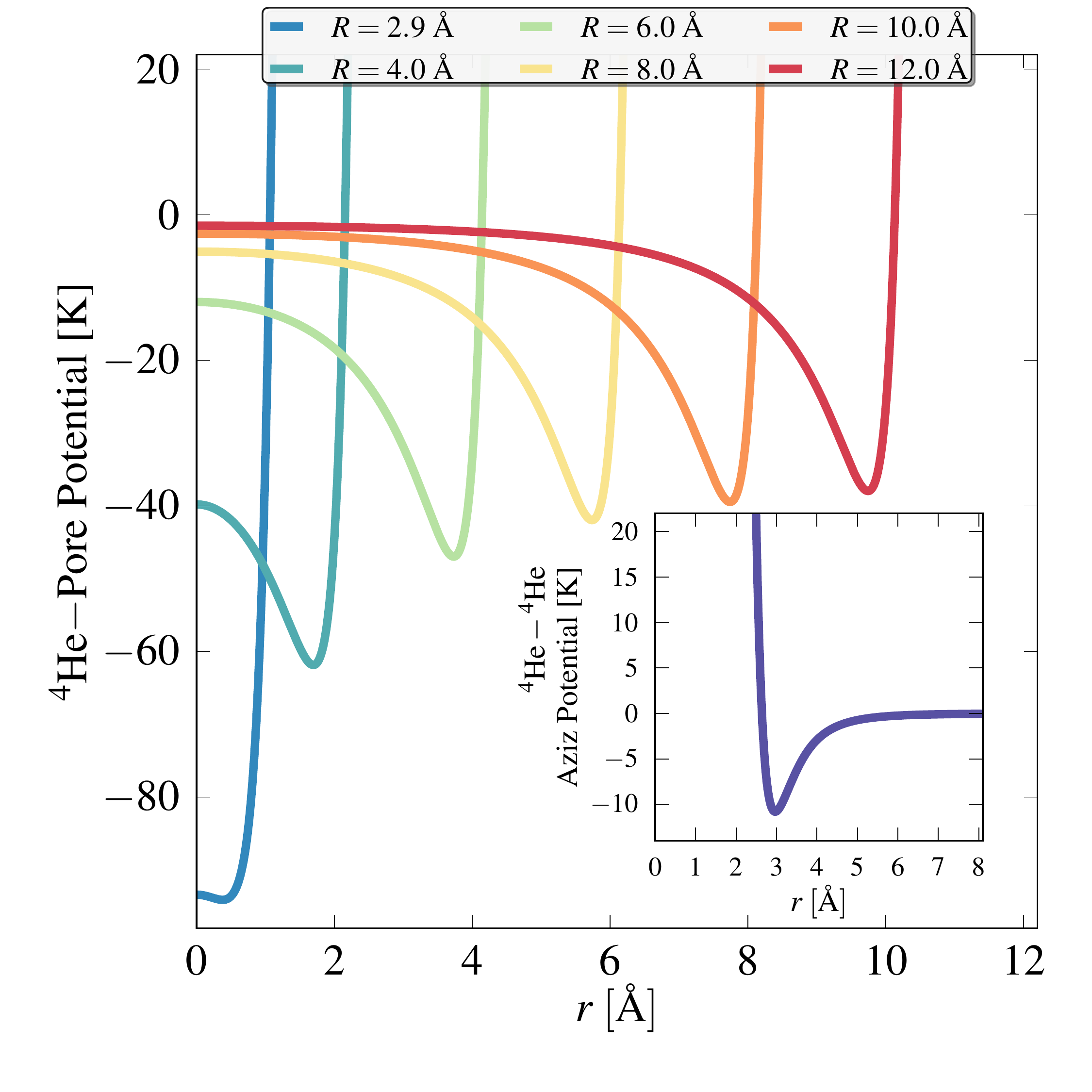}
\caption{The interaction potential $V$ between helium and the
walls of the nanopore constructed of amorphous Si$_3$N$_4$ for various radii
ranging from $R = 2.9\ang$ (left) to $R = 12.0\ang$ (right) calculated via
Eq.~(\ref{eqPorePotential}) where $r$ is the radial distance
of a helium atom from the axis of the cylinder.  The inset shows the Aziz
interaction potential $v$, described in Eq.~(\ref{eqAzizPotential}) for two helium
atoms separated by a distance $r$.}
\label{figPotential}
\end{figure}

We wish to perform WA simulations on a realistic physical model of the
experimental single nanopore geometry discussed in the introduction with
details in References~\onlinecite{Savard:2009eq} and \onlinecite{Savard:2011fe} and
thus must consider interactions between helium atoms, as well as the
interactions between helium and the atoms composing the material surrounding
the nanopore.  Our starting point is the general microscopic many-body
Hamiltonian
\begin{equation}
\hat{H} = \sum_{i=1}^{N}\left[- \frac{1}{2m}\hat{\vec{\nabla}}_i^2 +
\hat{V}(\vec{r_i}) \right] +
\sum_{i<j}\hat{v}(\vec{r}_i - \vec{r}_j)
\label{eqHamMicroscopic}
\end{equation}
where $\vec{r}_i$ are the spatial positions of the $N$ helium atoms and we
have set $\hbar = k_{\mathrm{B}} = 1$.  The natural units of length in the
nanopore systems are angstroms and in these reduced units, the mass of the
helium atoms is $m = 0.0826~\text{\AA}^{-2}\mathrm{K}^{-1}$. The interaction
energy between two helium atoms $v$, is given by the 1979 Aziz
potential\cite{Aziz:1979hs}
\begin{align}
\label{eqAzizPotential}
v(r) & = \epsilon u\left(\frac{r}{r_{\text{min}}}\right), \\
u(x) & = A\mathrm{e}^{-\alpha x} - \left(\frac{C_6}{x^6} + \frac{C_8}{x^8} +
\frac{C_{10}}{x^{10}} \right) \left \{
\begin{array}{rcl}
\mathrm{e}^{-(D/x-1)^2} & ; & x < D \\
1 & ; & x \ge D
\end{array}
\right.
\end{align}
with a set of parameters inferred from both first principles calculations
and experimental measurements to be $r_{\text{min}} = 2.9673\ang$, $A =
5.449\times10^5$, $\epsilon/k_{\mathrm{B}} = 10.8$~K, $\alpha = 13.353$,
$D=1.2413$, $C_6 = 1.3732$, $C_8 = 0.42538$ and $C_{10} = 0.1781$. This
potential, shown as an inset in Figure~\ref{figPotential}, 
%
%
contains both hardcore repulsion for distances less than $r_{\mathrm{hc}}
\simeq 2.64\ang$, a minimum at $r_{\text{min}}$ and weak attraction at large
separations.  The external potential, $V$ can be computed by modeling the
nanopore as long cylindrical cavity carved inside a continuous medium.  The
potential energy of interaction of a single helium atom with the atoms of the
medium can be obtained by integrating a Lennard-Jones pair potential as
described in Reference~\onlinecite{Tjatjopoulos:1988tn} yielding
\begin{equation}
V(r;R) = \frac{\pi \varepsilon\sigma^3 n}{3} 
\left[\frac{7}{32}\left(\frac{\sigma}{R}\right)^9 u_9\left(\frac{r}{R}\right)
-\left(\frac{\sigma}{R}\right)^3 u_3\left(\frac{r}{R}\right) \right]
\label{eqPorePotential}
\end{equation}
for a perfect cylindrical cavity of radius $R$ where $n$ is the number density
and $\sigma$ and $\varepsilon$ are the Lennard-Jones parameters of the
surrounding medium.  The functional coefficients are given by
\begin{align}
u_9(x) & = \frac{1}{240(1-x^2)^9} \left[(1091 + 11516x^2 + 16434x^4 + 4052x^6 + 35x^8)E(x^2)
\right. \nonumber \\
& \qquad \left. - 8(1-x^2)(1+7x^2)(97+134x^2+25x^4)K(x^2)\right] \\
u_3(x) & = \frac{2}{(1-x^2)^3} \left[(7+x^2)E(x^2) - 4(1-x^2)K(x^2)\right]
\end{align}
where $K(x)$ and $E(x)$ are the complete elliptical integrals of the first and
second kind respectively and $r$ is the radial distance from the axis of the
pore. Using the Lorentz-Berthelot mixing rules for the microscopic
Lennard-Jones parameters of amorphous
Si$_3$N$_4$\cite{Wendel:1992jq,Chakravarty:1997dy,Ching:1998jj} we set
$\varepsilon = 10.22$~K and $\sigma= 2.628\ang$  with $n =
0.078~\text{\AA}^{-3}$.  A plot of the pore potential for different radii is
shown in Figure~\ref{figPotential}.  The use of parameters for different media
such as silica glass\cite{Boninsegni:2010fg} will alter the overall energy
scale and may change the excluded volume experienced by atoms inside the pore
but will not qualitatively affect the findings of this study.

\section{Helium-4 Inside Nanopores}

Using this model for the interaction and confinement potentials, we have
computed the low temperature thermodynamics of fluid $^4$He inside nanopores
with lengths between $L=50$ and $200\ang$ and temperatures ranging from $T=0.5 -
2.0$~K using the WA.  We imagine the nanopore to be immersed in an essentially
infinite bath of helium maintained at saturated vapor pressure (SVP) which sets
the chemical potential in our grand canonical simulations to be $\mu = -7.2$~K.
A wide range of cylindrical pores with radii $R$ between $2.9$ and $12.0\ang$
have been considered where we assume periodic boundary conditions along the axis
of the cylinder.

\begin{figure}[t]
\includegraphics[width=\figwidth]{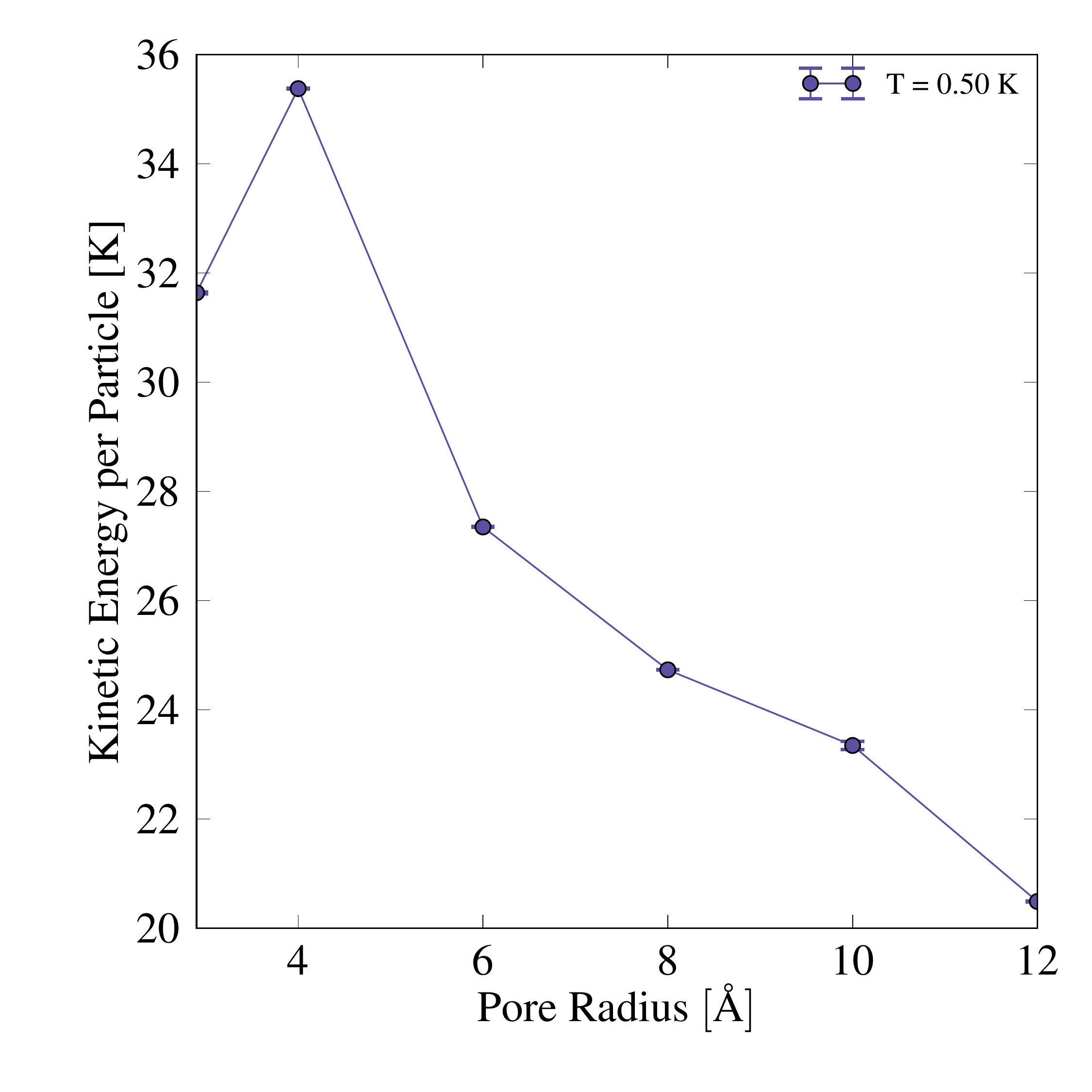}
\caption{\label{figKineticEnergy}The average kinetic energy per particle vs.
pore radius measured via a thermodynamic estimator at low temperature ($T =
0.5$~K) for a pore of length $L=100\ang$ held at $\mu=-7.2$~K.}
\end{figure}

We restrict the temperature in our simulations to be less than $T_\lambda \simeq
2.17$~K and wish to confirm that we are studying a \emph{low energy} quantum
regime.  This is most straightforwardly determined by measuring the average
kinetic energy per particle in our quantum Monte Carlo (QMC) simulations as
shown in Figure~\ref{figKineticEnergy} and comparing with the temperature.  
%
%
We observe a general trend of decreasing kinetic energy, from more than $30$~K
per particle for $R=2.9\ang$ to $20$~K per particle for $R=12\ang$.  The
non-monotonic behavior observed between $R=2.9\ang$ and $R=4.0\ang$ can be
attributed to the complete transverse confinement of helium in the narrowest
pore.  The minimum value of the kinetic energy observed is an order of magnitude
larger than all temperatures considered and we thus conclude our simulations
are dominated by quantum effects.  Unless otherwise specified, we fix the length
of the nanopore to be $L = 100\ang$ allowing three dimensional number densities
to be easily converted into particle numbers (usually near $1000$ helium
atoms).  The errorbars in Figure~\ref{figKineticEnergy} are the result of a
bootstrapping analysis of over $10^6$ individual measurements of the kinetic
energy using a thermodynamic estimator\cite{Jang:2001cl}.  We have fixed the
number of time slices (the discretization of the imaginary time interval
corresponding to the inverse temperature) to be $125$ per degree kelvin.
Possible Trotter error introduced by using a finite imaginary time step $\Delta
\tau$ has been analyzed and we
find that it is well described by a term proportional to $(\Delta \tau)^4$
consistent with the use of a fourth order path integral factorization of the
density matrix\cite{Jang:2001cl}. 

Assured we are in a low energy regime over the entire range of temperatures
considered in this study, we now switch our attention to the structures formed
by helium atoms inside the pore.  There exists a large literature on the types
of phases that are formed when confining a quantum fluid of helium in a
cylindrical cavity and a complex phase diagram has been predicted for carbon
nanotubes\cite{Stan:1998du,Gatica:2000jc,Cole:2000km,Gordillo:2000by,Gordillo:2007bt,Gordillo:2009hq}
and both smooth\cite{Urban:2006gh,Hernandez:2010dk,DelMaestro:2011dh} and
porous nanopores\cite{Rossi:2005hp,Rossi:2006cj}.  Advances owing to the
development of the WA have allowed this and a previous
work\cite{DelMaestro:2011dh} to study considerably larger radius pores at high
density and finite temperature in the grand canonical ensemble that approach
those studied in recent experiments\cite{Savard:2009eq,Savard:2011fe}. 

\begin{figure}[th]
\includegraphics[width=\figwidth]{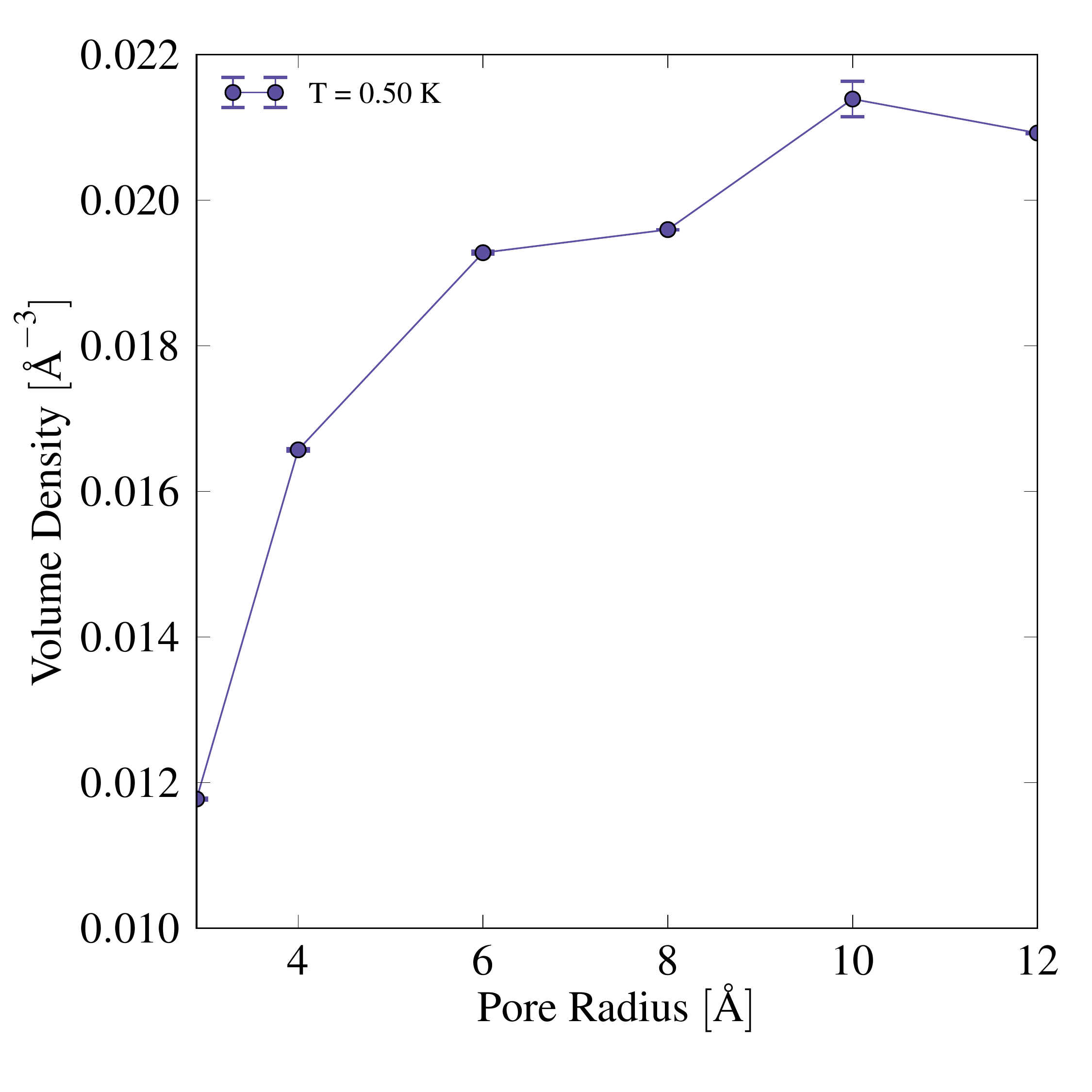}
\caption{\label{figVolumeDensity}The average volume density as a function of
nanopore radius for helium at saturated vapor pressure and $T=0.5$~K.}
\end{figure}

\begin{figure}
\includegraphics[width=0.9\textwidth]{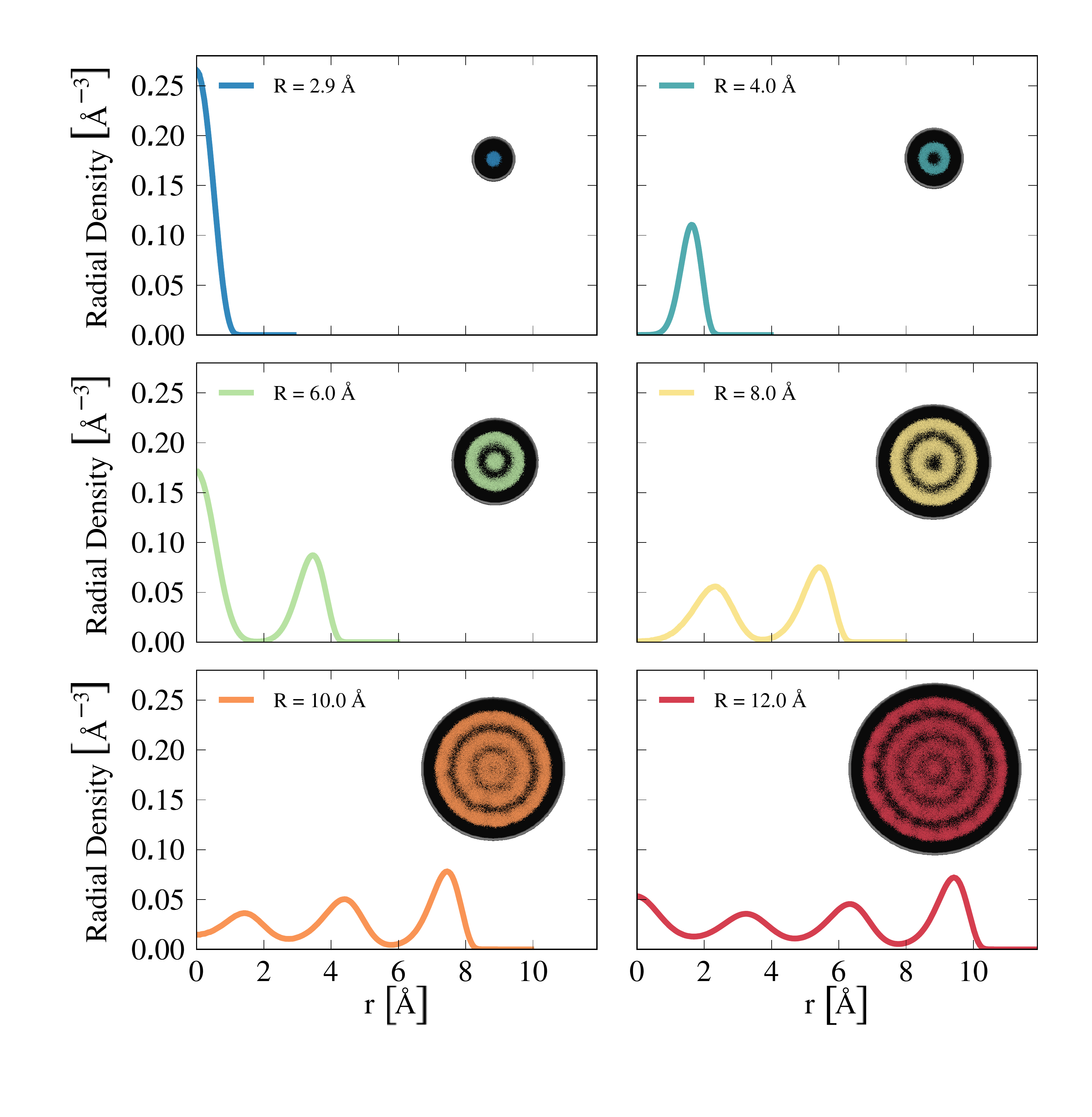}
\caption{\label{figRadialDensityArray} The radial number density $\rho_R(r)$ of
helium inside nanopores with radii $R = 2.9, 4.0, 6.0, 8.0, 10.0, 12.0\ang$ with
an inset showing a full instantaneous worldline configuration inside the pore
measured in the quantum Monte Carlo simulations projected on $z = 0$.  All pores
of $L=100\ang$ and are held at saturated vapor pressure corresponding to
$\mu=-7.2$~K.}
\end{figure}

For fixed chemical potential $\mu=-7.2$~K the average volume density  $\rho_{V}
= \langle N \rangle / \pi R^2 L$ inside the nanopore is shown in
Figure~\ref{figVolumeDensity}.
%
%
where $\langle \cdots \rangle$ indicates a thermodynamic average performed in
the quantum Monte Carlo.  The observed general trend is an increase in the
volume density as a function of radius continuing to a limiting value for large
$R$ that is approaching the expected bulk density of
$0.02198\text{\AA}^{-3}$\cite{Boninsegni:2006gc}.  The resulting phases of
helium inside the nanopore can be elucidated by measuring the radial density of
particles, $\rho_R(r)$ defined to be the number of particles found a distance
$r$ from the axis of a pore of radius $R$ normalized such that the linear
density $\rho_L$ is 
\begin{equation}
\rho_L \equiv \frac{\langle N \rangle}{L} = 2\pi \int dr r \rho_R(r)
\label{eqRadialDensity}
\end{equation}
where $\rho_R(r)$ includes an implicit average over the axial and angular
coordinates.  The results, shown in Figure~\ref{figRadialDensityArray} display a
progression of structures including a nearly one dimensional chain of helium
atoms for $R=2.9\ang$ to a series of three concentric shells surrounding a chain
for $R = 12.0\ang$.  
%
%
The insets display a snapshot of the discretized helium worldline configuration
(space and imaginary time coordinates) from the QMC simulations projected into
the $xy$-plane at $z=0$.

The observed oscillations in the radial density can be easily understood by
contemplating the filling of an empty nanopore through a classical analysis of
the potential interactions only (Figure~\ref{figPotential}).  Beginning with
the widest pores, we observe an excluded volume effect due to the \emph{hard}
Si$_3$N$_4$ wall, as well as a deep minimum of the potential near the surface.
This will naturally lead to an adsorption effect or wetting of the pore surface
with helium atoms forming a shell due to the cylindrical symmetry.  As helium
atoms continue to enter the pore, the adsorption will cause the area density of
the shell to increase until the average separation between atoms inside the
shell begins to approach the hard core radius $r_{\mathrm{hc}}$ of the
interaction potential.  At this point, it will become energetically favorable
to form a new layer inside the one adsorbed on the surface.  As this process
continues, a series of concentric cylindrical shells may be formed inside the
pore analogous to the layering observed in quantum films of
bosons\cite{Clements:1993jy}.

The results of the radial density for pores of different radii can be separated
into two natural groups based on the presence of helium at high linear density
near the center of the pore.   The existence of such an \emph{inner cylinder}
(IC) which can be thought of as a quasi-1D chain of helium atoms for a given
radius depends on the details of both the helium-pore and helium-helium
interaction potentials.  The pore wall (and thus the details of the wall
material, Si$_3$N$_4$ here) produces a region of excluded volume, forcing the
helium atoms to remain a distance greater than $r_e \simeq 1.32\ang$ from the
walls of the pore.  This sets the location of the outermost shell.  The
separation between shells is then restricted to be near the minimum of the Aziz
potential $r_{\mathrm{m}} \sim 3\ang$ with some weak dependence on the radius
of the pore due to screening of the inner shells.  We find that an IC exists
whenever the pore radius is approximately divisible by three: $R=2.9,6.0$ and
$12.0\ang$.

The presence of an IC of helium is intriguing, as we expect that such a
quasi-1d bosonic system should lack any long range order down to zero
temperature and be described by Luttinger liquid theory.  In the next section
we test this prediction by introducing the universal theory for an effective
harmonic fluid and use it to exhaustively study nanopores with radii $R =
2.9\ang$ and $R=12.0\ang$.

\section{A Luttinger Liquid Core}

In an attempt to understand the relevant low energy degrees of freedom for
quasi-1D helium atoms inside the pore, we begin by studying a microscopic Hamiltonian
like the one in Eq.~(\ref{eqHamMicroscopic}) in second quantized form for a
strictly one dimensional system of bosons
\begin{equation}
H = \int_0^L d z 
    \left [ 
        \frac{1}{2m}\partial_z \Psi^\dag(z) \partial_z \Psi(z)
        + \frac{1}{2}\int_0^L d z' \rho(z) v_{\text{1d}}(z-z') \rho(z) 
    \right ]
\label{eqHam2Q}
\end{equation}
where $\Psi^{\dag}(z) = \sqrt{\rho(z)}\mathrm{e}^{-i\phi(z)}$ is a bosonic
creation operator with $[\Psi(z),\Psi^\dag(z')] = \delta(z-z')$ and the
1d density and phase operators satisfy 
\begin{equation}
[\rho(z),\mathrm{e}^{i\phi(z')}] = \delta(z-z')\mathrm{e}^{i\phi(z)}.
\label{eqComRhoPhi}
\end{equation}
It is the ability to simulate such microscopic Hamiltonians over a range of
energy scales that makes the WA so attractive.   The stochastically exact
equilibrium properties of a system of strongly interacting bosons at finite
temperature can be determined using only fundamental input parameters such as
the particle mass $m$ and the details of the interaction potential. 

%
%
\subsection{Luttinger Liquid Theory}
\label{subsecLLT}

The manipulations (generally referred to as bosonization) that transform a
microscopic one dimensional interacting Hamiltonian like Eq.~(\ref{eqHam2Q})
into a universal description of the linear hydrodynamics of a quantum fluid
(either bosonic or fermionic) are by now
standard\cite{Haldane:1981gd,Giamarchi:2004qp,Cazalilla:2011dm,Cazalilla:2004dd}.
In particular, for bosonic systems, Reference~\onlinecite{Cazalilla:2004dd}
provides a detailed and pedagogical derivation of the Luttinger Liquid
Hamiltonian defined
by
\begin{equation}
H_{\text{LL}}-\mu N = 
    \frac{1}{2\pi}\int_0^L d z \left[v_J \left(\partial_z
            \phi\right)^2 + v_N \left(\partial_z \theta\right)^2\right],
\label{eqLLHam}
\end{equation}
where we have included only those terms which are formally relevant in the
renormalization group sense.  The velocities $v_J$ and $v_N$ are fixed by the
microscopic details of the underlying high energy model and the angular field
$\theta(z)$ appears through the redefinition of the density operator
\begin{equation}
\rho(z) \equiv 
    \left[ 
        \rho_0 + \frac{1}{\pi}\partial_z \theta(z)
    \right] 
    \sum_{m=-\infty}^{\infty} \mathrm{e}^{i 2 m \theta(z)}
\label{eqThetaDensity}
\end{equation}
where $\rho_0$ is the number density at $T=0$ in the thermodynamic limit.  The
underlying bosonic symmetry requires that Eq.~(\ref{eqThetaDensity}) in
conjunction with Eq.~(\ref{eqComRhoPhi}) produces the following commutation
relation
\begin{equation}
[\partial_z \theta(z),\phi(z')] = i \pi \delta(z-z').
\label{eqComThetaPhi}
\end{equation}
If the system described by Eq.~(\ref{eqLLHam}) exhibits Galilean invariance we
can identify
\begin{align}
v_J &= \frac{\pi\rho_0}{m}, \\
v_N &= \frac{1}{\pi \rho_0^2 \kappa}
\label{eqLLVelocities}
\end{align}
where $\kappa$ is the adiabatic compressibility in the limit $L\to \infty$,
$T\to 0$\cite{Haldane:1981gd}.  The connection of these effective velocities to
the underlying microscopic details of the original interacting Hamiltonian is
now clear: a highly incompressible state with $v_N \gg 1$ should have a nearly
constant density and thus $\partial_z \theta(z) \sim 0$ whereas a state
displaying phase coherence has $\partial_z \phi(z) \sim 0$ and thus at $T=0$
there should be a finite superfluid fraction with $v_J \gg 1$.

For a system with periodic boundary conditions
in the $z$-direction, our original boson field must satisfy $\Psi^\dag(z + L) =
\Psi^\dag(z)$ leading to a mode expansion of $\theta(z)$ and $\phi(z)$ indexed
by wavevector $q = 2\pi n /L$ where $n$ is an integer\cite{Haldane:1981gd}
\begin{align}
\theta (z) &= \theta_0 + \frac{\pi z}{L} (N-N_0)
-i\left(\frac{v_J}{v_N}\right)^{1/4} \sum_{q\neq 0} 
\left \lvert \frac{\pi}{2 q L}\right \rvert^{1/2} 
\mathrm{e}^{i q z}(b_q^\dagger +b^{\phantom\dag}_{-q})\mathrm{sgn}(q), 
\label{eqTheta} \\
\phi (z) &= \phi_0+\frac{\pi J
z}{L}-i\left(\frac{v_J}{v_N}\right)^{-1/4}\sum_{q\neq 0}
\left \lvert \frac{\pi}{2 q L}\right \rvert^{1/2} \mathrm{e}^{i q z}(b_q^\dagger
-b^{\phantom\dag}_{-q}).  
\label{eqPhi}
\end{align}
Here, $b^\dag_q\, (b_q)$ is a bosonic creation (annihilation) operator for modes
corresponding to long wavelength density fluctuations and the operator $J$ has
even integer eigenvalues indexing the topological winding number of the phase
field $\phi(z)$, $N_0 = \rho_0 L$ and $N$ is the boson number operator.
Substituting Eqs.~(\ref{eqTheta}) and (\ref{eqPhi}) into (\ref{eqLLHam})
yields the mode expanded Hamiltonian
\begin{equation}
H_{\text{LL}} - \mu N = \frac{\pi v}{2 K L}J^2 + \frac{\pi v K }{2 L}
(N-N_0)^2 + \sum_{n\ne0}v|q|b^\dag_q b_q ,
\label{eqHamLLModeExpansion}
\end{equation}
where we have dropped a non-universal constant and defined a new velocity $v =
\sqrt{v_J v_N}$ measuring the famous linear dispersion of the low energy
density modes.  We have introduced the Luttinger parameter  $K =
\sqrt{v_N/v_J}$\cite{DelMaestro:LLNote} and although it is well known that no
quantum phase transition can occur as a function of interactions in a truly 1d
system described by Eq.~(\ref{eqHamLLModeExpansion}), $K$ can be tuned to
initiate a $T = 0$ crossover between a state with algebraic density wave order
at $K = \infty$ to one with quasi-long range superfluid correlations at $K=0$.

The advantages of having the Hamiltonian in this form are unmistakable due to
its quadratic nature and the ease with which we can compute averages in the
oscillator basis. For example, one can exactly determine the grand partition
function  $\mathcal{Z} = \text{Tr}\, \exp[-(H_{LL} - \mu N)/T]$ in terms of
known special functions\cite{DelMaestro:2010cz} allowing for the straightforward
(although possibly tedious) calculation of all two body correlation functions
and thermodynamic observables in terms of the temperature $T$, system size $L$,
Luttinger velocity $v$ and  Luttinger parameter $K$.  We will focus on the
derivation of a single observable, the density-density correlation function
$\langle \rho(z) \rho(0) \rangle$. Such density correlations are of great
interest, as they are easily measured in numerical simulations and are related
via a Fourier transform to the experimentally measurable structure factor.  In
addition, their form provides an intuitive qualitative picture of the types of
fluctuations (phase or density) which are dominant. This knowledge can help to
pinpoint which region of the 1d crossover phase diagram a given system resides
in.  Starting from the definition of the density operator in
Eq.~(\ref{eqThetaDensity}) we have (keeping only the slowest decaying terms) 
\begin{equation}
\langle \rho(z) \rho(0) \rangle \approx \rho_0^2 + \frac{1}{\pi^2}\langle \partial_z
\theta(z) \partial_z \theta(0) \rangle + 2 \rho_0^2 \langle \mathrm{e}^{2 i
\theta(z)} \mathrm{e}^{-2 i \theta(0)}\rangle.
\label{eqPCFDef}
\end{equation}
Next, using the mode expansion for $\theta(z)$ in Eq.~(\ref{eqTheta}) the expectation
values can be computed to give\cite{DelMaestro:2011dh}
\begin{multline}
\langle \rho (z)\rho (0) \rangle = \rho_0^2 + \frac{1}{2\pi^2 K} \frac{d^2}{dz^2}
\ln \theta_1 \left[\frac{\pi z}{L},\mathrm{e}^{-\pi v/LT} \right] \\
+ \mathcal{A} \cos (2\pi \rho_0 z) \left\{
    \frac{2\,\eta \left(\frac{i v}{LT}\right) \mathrm{e}^{-\pi v/ 6LT}}
{\theta_1\left(\frac{\pi z}{L},\mathrm{e}^{-\pi v/LT}\right)}\right\}^{2/K}
\label{eqPCF} 
\end{multline}
where $\theta_1(x,y)$ and $\eta(i s)$ are the first Elliptical Theta function
and Dedekind Eta function respectively.  $\mathcal{A}$ is a non-universal
constant dependent on the short-distance properties of the system.  Although
Eq.~(\ref{eqPCF}) may appear daunting at first glance, in the thermodynamic
limit $LT/v \to \infty$, it simplifies to\cite{Haldane:1981gd}
\begin{equation}
\langle \rho(z) \rho(0) \rangle \to \rho_0^2
- \frac{1}{2\pi^2 K z^2} + \frac{\mathcal{A}}{z^{2/K}}\cos(2\pi \rho_0 z)
\label{eqPCFThermoLimit}
\end{equation}
where it is now clear that $K \gg 1$ corresponds to a strong tendency towards
density wave order.

We are now in a position to test the prediction that for some special radii,
the core of a helium-4 filled nanopore can be fully described by Luttinger
liquid theory.   We can directly compare the pair correlation function measured
in the quantum Monte Carlo with the Luttinger liquid prediction of
Eq.~(\ref{eqPCF}).   Before doing so, let us make a brief digression to study
the connection between the interactions in the truly one dimensional
Hamiltonian in Eq.~(\ref{eqHam2Q}) and those experienced by helium atoms inside
the quasi-one-dimensional environment inside the nanopore.

%
%
\subsection{Effective Interactions and Linear Density in the Inner Cylinder}

\begin{figure}
\includegraphics[width=\figwidth]{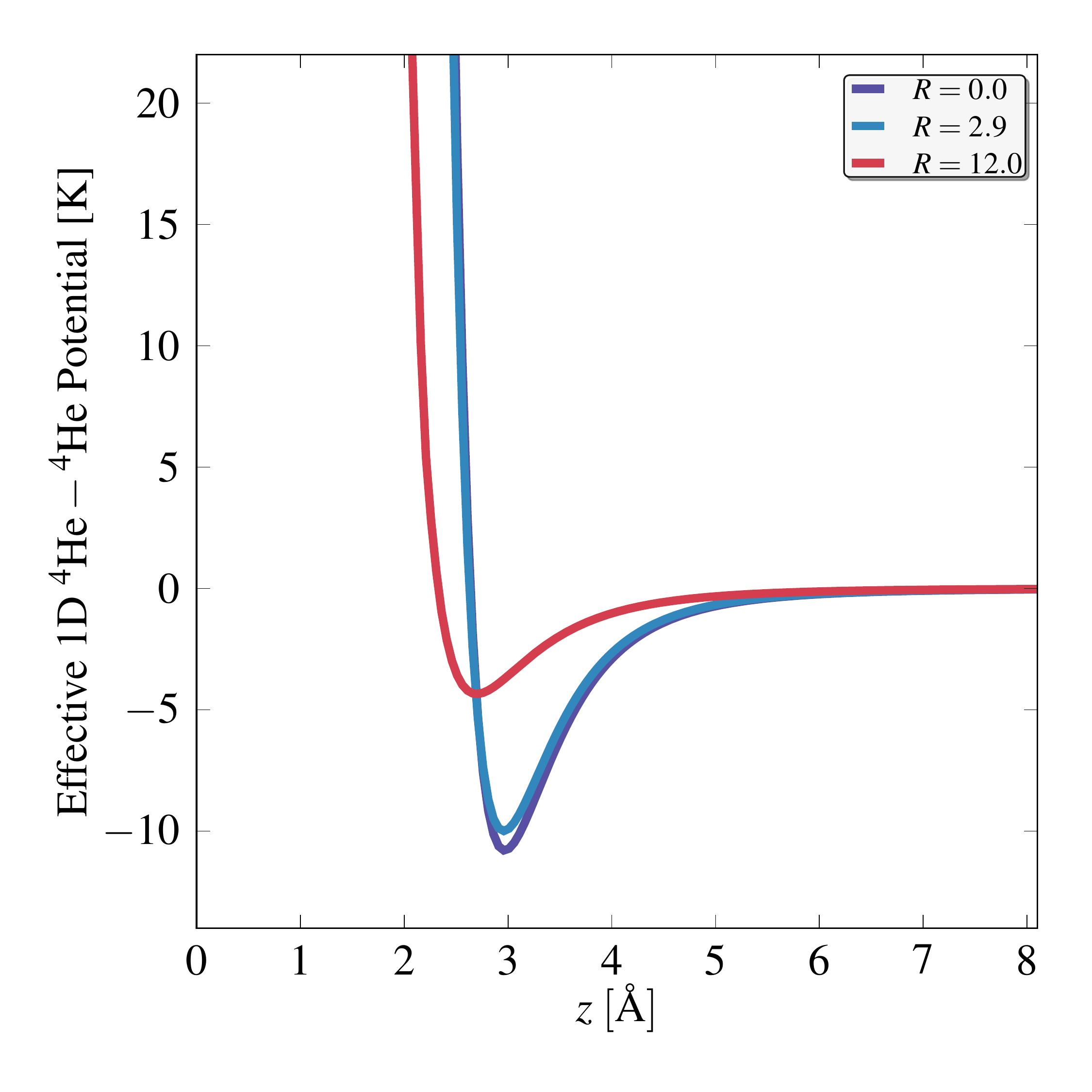}
\caption{\label{figEff1dPotential} The effective interaction energy between two
helium atoms in strictly one dimension (lowest curve) as well as for the
narrowest ($R = 2.9\ang$) and widest ($R=12.0\ang$) pores considered in this
study.  The radial density used in the numerical evaluation of
Eq.~(\ref{eqV1d}) was measured at $T = 0.5$~K, but the form of the potential is
only weakly dependent on temperature when $T < 1.0$~K.}
\end{figure}

\begin{figure}
\includegraphics[width=\figwidth]{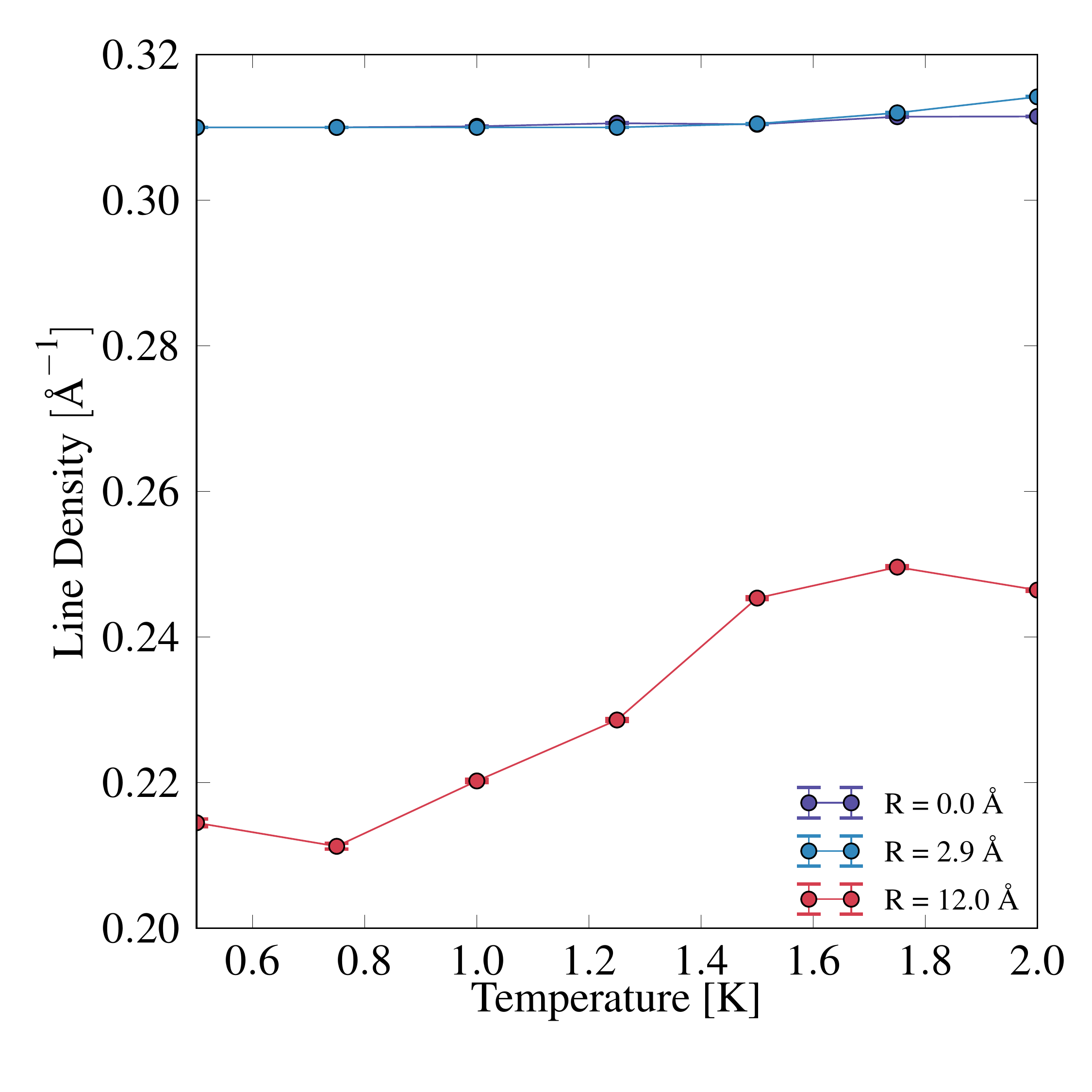}
\caption{\label{figLinearDensity} The number of particles per angstrom for
helium atoms in strictly one dimension (with $\mu = 85$~K), a pore with $R =
2.9\ang$ and inside the inner cylinder with $r < r_{\text{IC}} = 1.75\ang$ for
$R = 12.0\ang$. All pores have length $L=100\ang$ and the finite radius pores
are held at saturated vapor pressure with $\mu=-7.2$~K.}
\end{figure}

The full interaction and external potentials inside the nanopore defined in
Eqs.~(\ref{eqAzizPotential}) and (\ref{eqPorePotential}) and displayed in
Figure~\ref{figPCFR0} can be related to the 1d potential of Eq.~(\ref{eqHam2Q})
through 
\begin{equation}
v_{\text{1d}}(z) = \frac{1}{\rho_{L}^2} \int d^2 r \int d^2 r' v(\vec{r}
            -\vec{r}')\rho_{\mathrm{R}}(r) \rho_{\mathrm{R}}(r')
\label{eqV1d}
\end{equation}
where $\vec{r} = (r,\varphi,z)$ is a vector in cylindrical coordinates and
$\rho_{\mathrm{R}}(r)$ is the radial density defined in
Eq.~(\ref{eqRadialDensity}). The form of this effective interaction potential
can be investigated using the measured data from our QMC simulations with the
results shown in Figure~\ref{figEff1dPotential}.
%
%
We observe that the effective potential in the $R = 2.9\ang$ pore very closely
reproduces the interactions that would be experienced in a system of helium
atoms interacting in strictly one dimension ($R = 0$).  For the largest radius pore of
$R=12.0\ang$ however, screening from the concentric shells leads to a potential
with a much weaker minimum (about half the depth of the unscreened bulk Aziz
potential) with its location shifted to smaller separations.   The
effective 1d interactions in the axial direction experienced by atoms in the
center of the pore influences the average separation between atoms which can be
determined from the average linear density in Figure~\ref{figLinearDensity}.
%
%
We observe almost no temperature dependence for small radii and the agreement
of the linear density for $R=0$ and $R=2.9\ang$ is by design, with the chemical
potential having been tuned to $\mu_\text{SVP} - V(0;2.9) \simeq 85$~K for the
one dimensional ($R=0$) system to reproduce the energetic confinement displayed
in Figure~\ref{figPotential} at low temperature.  The numerical value of
$\rho_{0}$, defined as the zero temperature thermodynamic limit of the linear
density in the core is measured to be $\rho_{0,0} \simeq \rho_{0, 2.9} =
0.3100(1)\text{\AA}^{-1}$ (with the number in brackets indicating the
uncertainty in the final digit and a second subscript being the pore radius in
angstroms).  The fact that this value is slightly smaller than
$r_{\text{min}}^{-1}$ can be attributed to quantum effects.  For $R=12.0\ang$,
the observed density of $\rho_{0,12} = 0.2144(5)~\text{\AA}^{-1}$ is only weakly
related to the minimum of the effective 1d potential shown in
Figure~\ref{figEff1dPotential} as there is a temperature dependent amount of
exchange of helium atoms between the inner cylinder and surrounding shells.

%
%
\subsection{Density Correlations}

Having studied the effective one dimensional potential felt by helium atoms
inside the pore, and determined the linear density in the inner cylinder we are
now in a position to evaluate the efficacy of Eq.~(\ref{eqPCF}) in describing
density correlations in the nanopore.  A first glance at this expression
indicates that there are \emph{four} fitting parameters,
($\rho_0,v,K,\mathcal{A}$) to be determined; a number that would seemingly
allow large flexibility (and thus limited accuracy) of any least squares
fitting procedure.  However, one of the most attractive features of
Eq.~(\ref{eqPCF}) is that it provides a prediction for the full analytical form
of the finite size and temperature scaling behavior of $\langle \rho(z) \rho(0)
\rangle$, a rare luxury indeed.  We thus have a stringent procedure for
confirming Luttinger liquid behavior in the nanopore: (1) perform simulations
for different pore radii, lengths and temperatures measuring thermodynamic
estimators and correlation functions.  (2) The average linear density in
the central region of the pore can be determined as a function of
temperature and extrapolated for each $L$ and $R$ to $T=0$, this fixes
$\rho_{0,R}$. (3) For each radius which displays a finite density of helium
atoms at $r=0$, perform a least squares fit of the density-density correlation
function measured in the QMC to Eq.~(\ref{eqPCF}) at a single $L$ and $T =
0.5$~K (the lowest temperature measured).  This fixes the remaining three
parameters, $v_R$, $K_R$ and $\mathcal{A}_R$ where the subscript $R$ will be
used to distinguish results for different finite radius pores.  These parameters
are \emph{intrinsic} to the zero temperature thermodynamic limit of
Eq.~(\ref{eqLLHam}) and provided that helium-4 inside the core is behaving as a
LL, \emph{cannot} exhibit any temperature dependence.  (4) In other words, for a
given radius, a single fit to Eq.~(\ref{eqPCF}) at fixed temperature and system
size is enough to compute the LL prediction for the pair correlation function at
all other temperatures and sizes and we explore the resulting predictions for
$R=0,2.9$ and $12.0\ang$ below.  Obviously we are constrained by the low energy,
long wavelength region of applicability of $H_{\text{LL}}$ assumed throughout
this study and it is easily confirmed that deviations appear at short lengths
and high temperature.

\begin{figure}
\includegraphics[width=\figwidth]{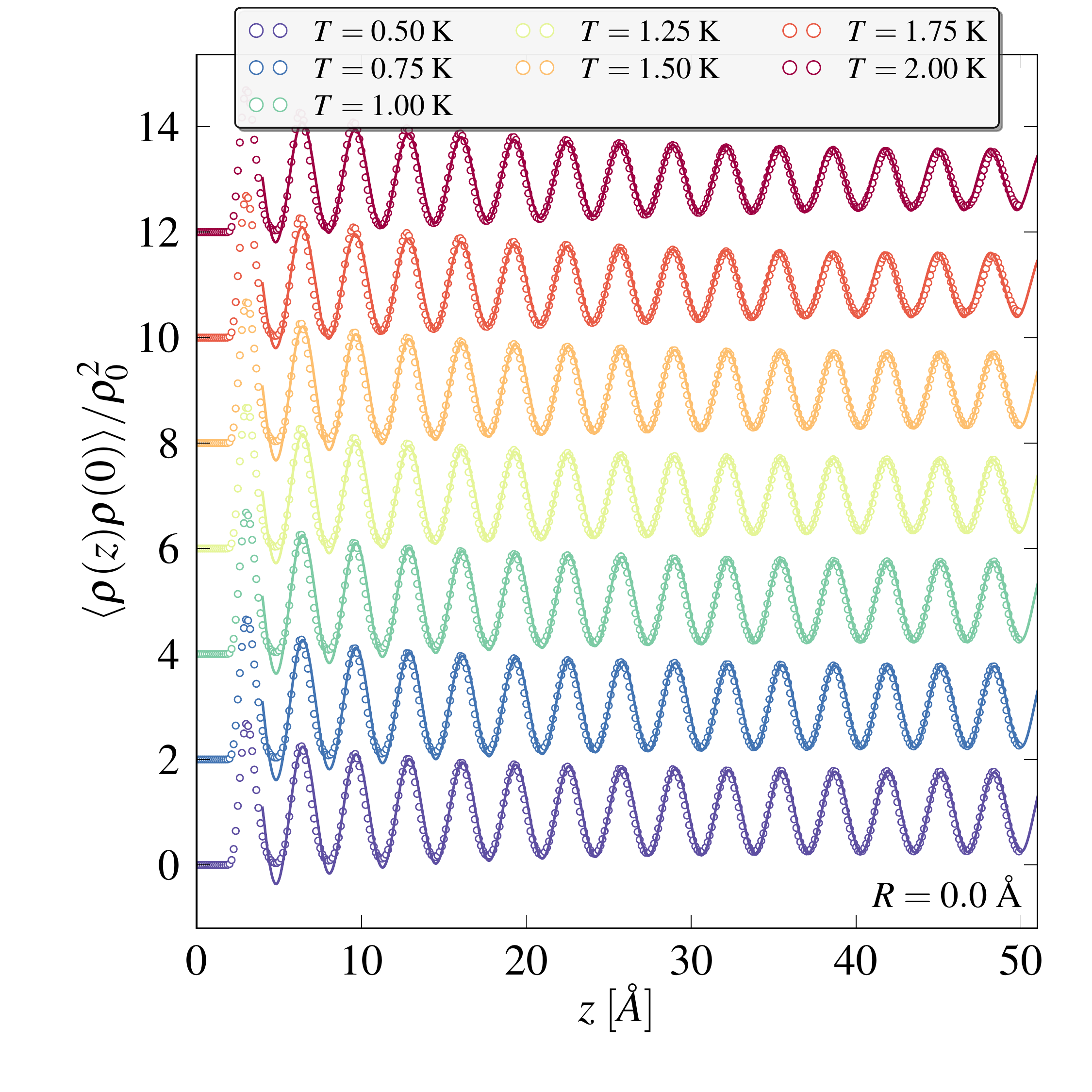}
\caption{\label{figPCFR0} Simulation data (circles) and a fit to Eq.~(\ref{eqPCF})
(lines) for the pair correlation function of strictly one dimensional helium
atoms ($R = 0$).  The chemical potential ($\mu = 85$~K) has been set to
approximate the energetically confining environment coming from the pore wall
(see text).  Error bars are smaller than the symbol size and data in the main
panel has been given a vertical $T$-dependent shift for clarity with $T=0.5$~K
at the bottom and $T = 2.0$~K at the top.}
\end{figure}

To orient ourselves we begin by studying the strictly one dimensional system,
which we fully expect to be well described by LL theory.  In
Figure~\ref{figPCFR0} we plot the results of QMC simulations for the
density-density or pair correlation function (symbols) for a chain of helium
atoms with $L = 100\ang$ and $\mu = 85$~K.  
%
%

\begin{figure}
\includegraphics[width=\figwidth]{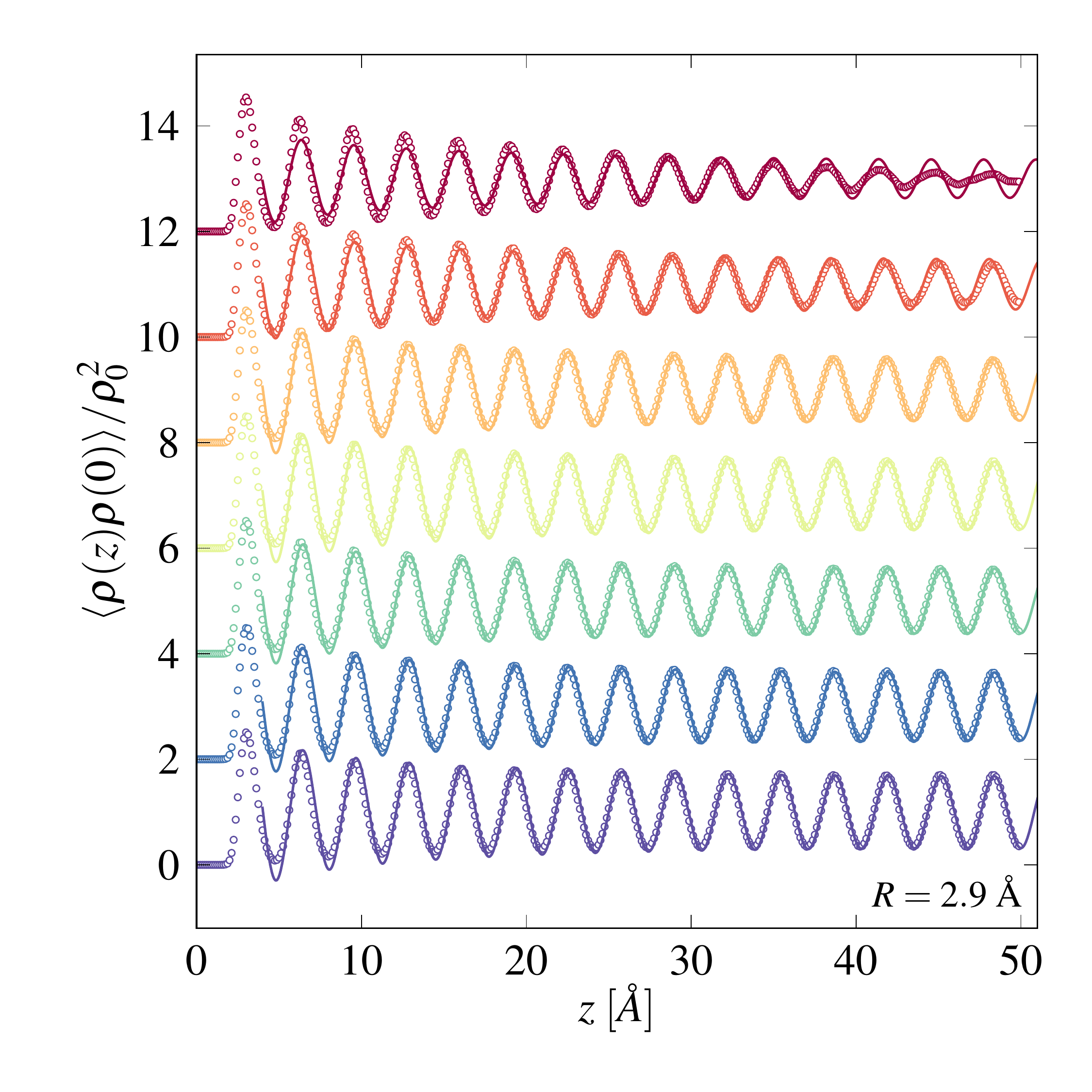}
\caption{\label{figPCFR3} Simulation data (symbols) and a fit to Eq.~(\ref{eqPCF})
(lines) for the pair correlation function along the axis of a nanopore with
$R=2.9\ang$ and $L = 100\ang$.  Different curves correspond to increasing
temperature from $T=0.5$~K (bottom) to $T=2.0$~K (top) (vertical shifts have
been added for clarity) with the legend displayed in Figure~7.}
\end{figure}

\begin{figure}
\includegraphics[width=\figwidth]{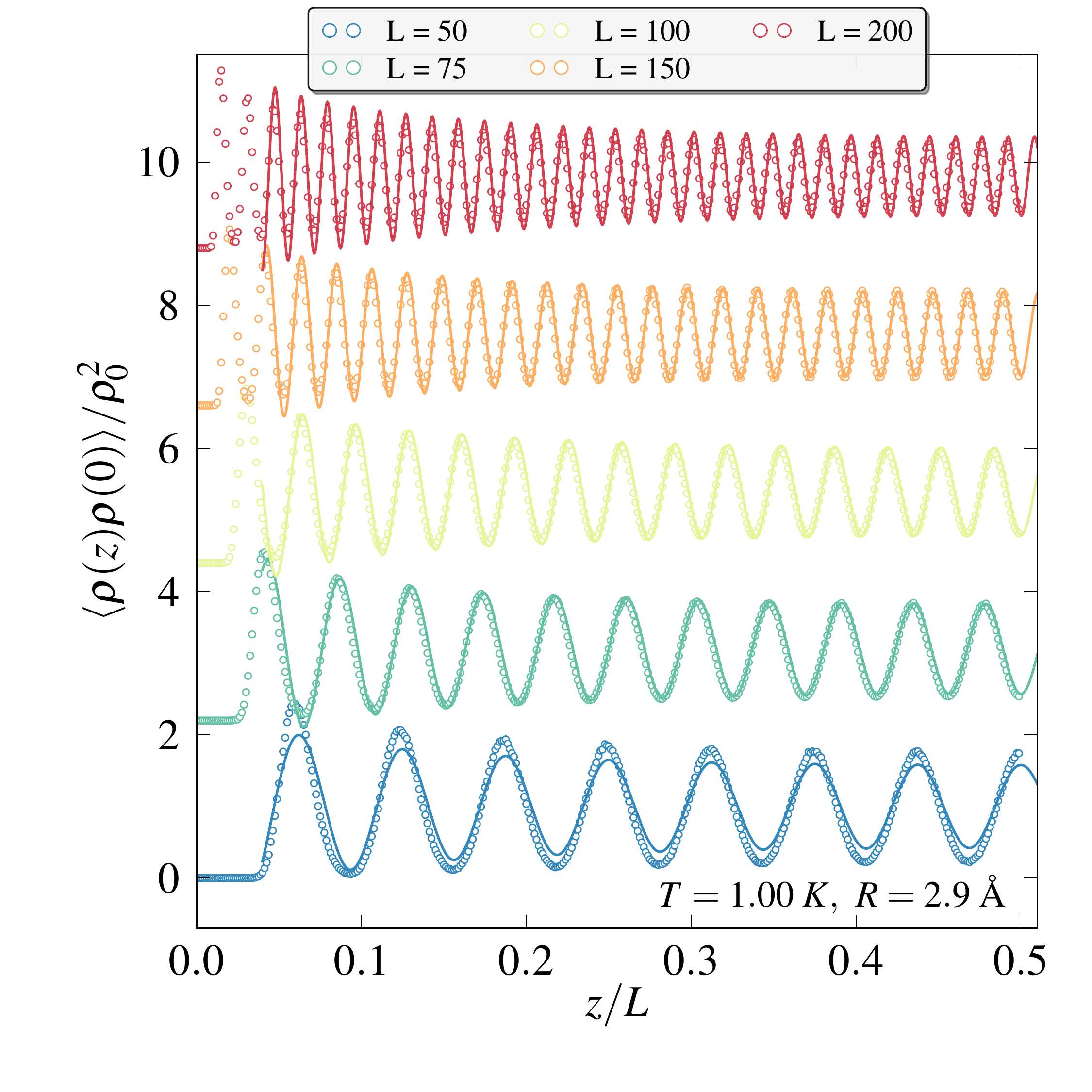}
\caption{\label{figPCFR3L} Simulation data (symbols) and a fit to Eq.~(\ref{eqPCF})
(lines) for the pair correlation function along the axis of a nanopore with $R
= 2.9\ang$ for various pore lengths $L$ (increasing from bottom to top).  
As in previous figures, a vertical shift has been added to distinguish the
curves and accentuate the scaling behavior. Note the use of a dimensionless
abscissa in order to plot spatial correlations for different pore lengths on
the same scale.}
\end{figure}

Performing the aforementioned fits at $T=0.5$~K, we find $v_0 =
75(1)\ang\text{K}$ and $K_0=6.4(3)$ where stochastic errors in QMC data produce
uncertainty in our regression procedure.  The exact value of $\mathcal{A_0}$ is
not relevant in the subsequent discussion and it will not be mentioned further.
Fixing all parameters, Eq.~(\ref{eqPCF}) was used to produce the solid lines in
Figure~(\ref{figPCFR0}) where a vertical shift has been included to allow for
the visual differentiation of the effects of temperature.  We observe spectacular
agreement over the entire range of temperatures considered, noting that no
additional fitting was required above $T = 0.5$~K.  We now perform an identical
procedure for the narrowest finite radius pore, $R=2.9\ang$, with the results
for the density-density correlation function shown in Figure~\ref{figPCFR3}.
%
%
Similar to the case of 1d helium $(R = 0)$, we observe persistent oscillations
out to the largest distances possible in the $L=100\ang$ pore with periodic boundary
conditions indicating a strong tendency towards density wave order.  As the
helium-pore interaction is independent of the axial coordinate $z$ and the radial
density for $R=2.9\ang$ (Figure~\ref{figRadialDensityArray}) exhibits only a
single central chain of atoms, Galilean invariance further restricts the ratio
$v_{2.9}/K_{2.9} = \pi \rho_{0,2.9} / m$\cite{Haldane:1981gd}.  Extracting values in the
presence of this constraint at $T = 0.5$~K yields $v_{2.9} = 70(3)\ang\text{K}$
and $K_{2.9} = 6.0(2)$, in close agreement with the 1d chain of helium atoms as
expected.  For the finite radius pore however, we begin to observe deviations
from LL predictions both at small distances and high temperature.  Once the
relevant thermal lengthscale, $\ell_T \sim v/T$ is on the order of the pore
diameter, the system can no longer be thought of as quasi-1d and we expect
significant corrections to arise from the thermal excitation of transverse
modes.  The values of the LL velocity and interaction strength determined in
this way can also be used to test the predicted finite size scaling of
Eq.~(\ref{eqPCF}) at fixed $T = 1.0$~K as seen in Figure~\ref{figPCFR3L}.
%
%
We stress that the nearly perfect agreement seen between simulation data and LL
theory in Figure~\ref{figPCFR3L} for $L \ge 75$ does not require any
additional fitting parameters and is a direct consequence of the predictive
power of the universal hydrodynamics of $H_{\text{LL}}$.

\begin{figure} 
\includegraphics[width=\figwidth]{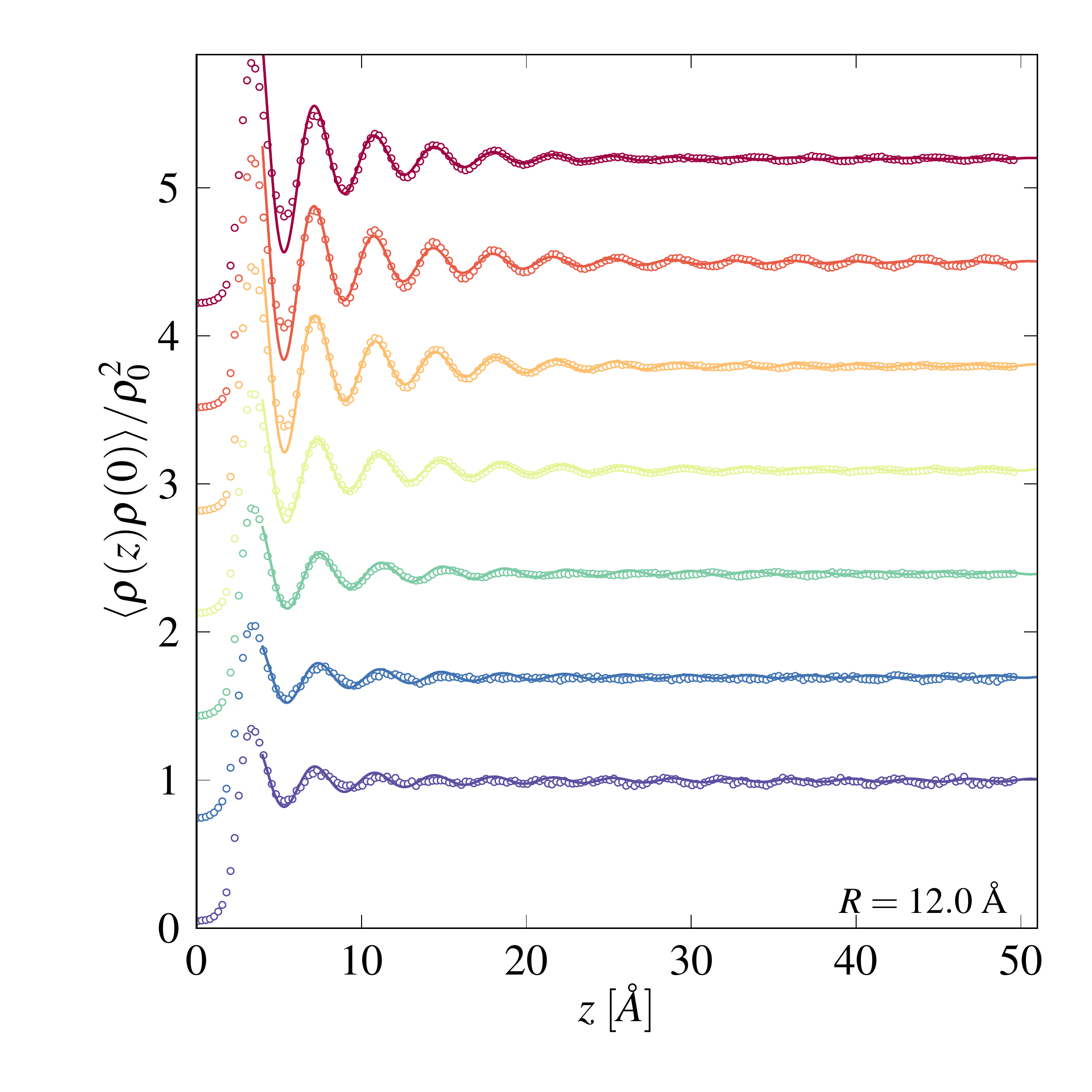}
\caption{\label{figPCFR12} Simulation data (symbols) and a fit to
Eq.~(\ref{eqPCF}) (lines) for the inner cylinder pair correlation function
along the axis of a nanopore with $L = 100\ang$ and $R = 12.0\ang$ for helium
atoms with $r < r_{\text{IC}} = 1.75\ang$.  The curves have been shifted for
clarity and correspond to increasing temperature from $T = 0.5$~K (bottom) to
top $T=2.0$~K (top) with with the legend displayed in Figure~7.}
\end{figure}

Shifting attention to the $R=12.0\ang$ pore, in order to make an effective
comparison we will focus the analysis on only those helium atoms which
spontaneously find themselves inside the inner cylinder.  The precise
definition of which helium atoms are inside the IC is somewhat arbitrary, but
we define this to coincide with the location of the first minimum in the radial
density in the presence of finite density at $r=0$.  For the pores considered
here, this corresponds to $r_{\text{IC}} = 1.75\ang$.  We expect that the
presence of particle exchanges between the central chain of helium atoms and
the surrounding cylindrical shells should begin to play an important role in
the `melting' of the quasi-long range density wave order observed in narrow
pores.  The resulting axial pair correlation function measured in
our QMC simulations is shown in Figure~\ref{figPCFR12}. 
%
%
The data show density-density correlations in the inner cylinder that decay
much faster than those seen for $R=2.9\ang$.   The lack of Galilean invariance
due to atoms in the surrounding concentric shells adds some additional freedom
to the now familiar fitting procedure (and thus more uncertainty) and it has
been determined that $v_{12} = 42(2)\ang\text{K}$ and $K_{12} = 1.3(2)$.  This
value of the Luttinger parameters is nearly a factor of five smaller than that
found for the strictly one dimensional system and indicates that this nanopore
is in a region of phase space starting to be dominated by superfluid
fluctuations.  

The success of the Luttinger liquid prediction for $\langle \rho(z)\rho(0)
\rangle$ in Eq.~(\ref{eqPCF}) in describing the density-density correlations
measured in our quantum Monte Carlo simulations is convincing evidence that
a quantum fluid of helium-4 in pores with $R<3\ang$ can be described by LL
theory and that the central region of wider pores may display emergent harmonic
fluid-like properties. In the next section, we discuss the implications of
these results and argue that the determination of $K$ and $v$ for a given
system is not simply an academic exercise but has real consequences for the
stability of the Luttinger liquid phase.

\section{Discussion}

Although it appears that LL theory works exceedingly well in describing pair or
density-density correlations for the finite radius pores considered in this
study, it is natural to ask whether there is something special about this
particular observable.  As mentioned in Section~\ref{subsecLLT} the utility of
the harmonic fluid description of 1d systems is that the resulting mode expanded
Hamiltonian of Eq.~(\ref{eqHamLLModeExpansion}) is quadratic (neglecting
formally irrelevant operators) and thus the full grand partition function can be
computed in closed form.  We can therefore compare other two body correlation
functions computed within Luttinger liquid theory with those measured in the QMC
using the values of $v_R$ and $K_R$ determined in this study and displayed in
Table~\ref{tabLL}.
%
\begin{table}
\caption{\label{tabLL} Values for the Luttinger velocity and parameter for
helium atoms in 1d and in the quasi-1d core of nanopores by comparing the
Luttinger liquid prediction with quantum Monte Carlo measurements of the pair
correlation function.  The number in brackets indicates the uncertainty in the
final digit.} 
\begin{ruledtabular}
\begin{tabular}{ccc} 
   Radius $R$ [\AA]  & Luttinger Velocity $v$ [\AA K] & Luttinger Parameter $K$ \\
   \hline \\[-4ex]
   0.0 & 75(1) & 6.4(3) \\ 
   2.9 & 70(3) & 6.0(2) \\ 
   12.0 & 42(2) & 1.3(1) \\
\end{tabular} 
\end{ruledtabular}
\end{table}
%
We have measured the axial one body matrix $n(z) = \langle
\Psi^\dag(z)\Psi(0)\rangle$ in the QMC\cite{Boninsegni:2006gc} and find
acceptable agreement with LL theory (with no new fitting parameters) at low $T$.
The large value of $K \approx 6$ for $R = 0$ and $R=2.9\ang$ adds some
complications when comparing numerical results for some scalar (non-correlation
function) observables with predictions coming from bosonization.   For example,
the strong tendency towards density wave order $\partial_z \theta(z) \sim 0$
displayed in Figures~\ref{figPCFR0}-\ref{figPCFR3} points to a compressibility
$a\kappa \sim \langle N^2 \rangle - \langle N \rangle^2 \approx 0$ and
thus an analysis of the probability distribution function for particle number
fluctuations like the one performed in Reference~\onlinecite{DelMaestro:2010cz} is
not feasible.  The conjugate relationship between density and phase variables in
the LL theory (Eq.~(\ref{eqComThetaPhi})) would predict that the boson phase
$\phi(z)$ should be totally disordered resulting in a vanishing superfluid
density which is observed in the numerics.

For the $R=12\ang$ pore, the analysis presented in the previous section is
based on a fraction of the total number of ${}^4$He atoms in the pore, those
that dynamically have $r < r_{\text{IC}}$ and find themselves in the core
of the nanopore.  The justification for this originates in the idea that we may
be able to regard the pore as a coupled multi-component LL, with cylindrical
shells replacing the legs of previously studied ladders\cite{Orignac:1998ir}.
Guided by these results we assume that only a single gapless degree of freedom
may survive as a ``center of mass mode'' in the low energy effective field
theory, due to tunneling between the shells.  In the nanopore, this tunneling
has two origins corresponding to the physical mobility of particles between
shells as well as multi-particle quantum exchange cycles which may dynamically
connect them at short time scales.

With this in mind, let us re-analyze the slight discrepancies between the
quantum Monte Carlo data in Figure~\ref{figPCFR12} and LL theory.  At the
lowest temperature considered, $T=0.5$~K, the simulation data
appears to show oscillations with a period that is slightly larger
than that predicted by $1/(2\pi \rho_{0,12})$.  This is most likely attributed
to the physical exchange of particles between the inner cylinder and the surrounding
shells producing a greater uncertainty in $\rho_{0,12}$ than is reflected in
errorbars and making the extrapolation to zero temperature a difficult task.
The amplitude of the oscillations in the pair correlation function is also
slightly overestimated by LL theory at low temperature.  Again, the
finite radius of the pore is to blame, resulting in a finite superfluid fraction
of helium (as measured via the usual winding number
estimator\cite{Pollock:1987ta} along the axis of the pore) of $\rho_s/\rho_0
\sim 0.2$ at $T = 1.0$~K.  Transverse degrees of freedom that are not frozen
out at this temperature lead to an increase in multi-particle exchanges that
enhance superfluidity in the nanopore.  

The opposite behavior appears to occur at high temperature with the simulation
data showing weaker decay than predicted by LL theory with thermal fluctuations
being unable to fully quench the proclivity towards density wave order.  This
is the opposite effect observed for $R=2.9\ang$ in Figure~\ref{figPCFR3} and it
can possibly be explained through stabilization of the IC from attraction with
surrounding atoms.

In a previous work\cite{DelMaestro:2010cz}, finite size corrections to 
$\rho_{0,R}$ arising from the inclusion of higher order formally irrelevant
terms in the effective Hamiltonian of Eq.~(\ref{eqLLHam}) have been discussed
at great length. However, due to the relatively large energy scales at play in the
nanopore (Figure~\ref{figKineticEnergy}) they are less important here and are
not major players in any observed discrepancies between simulation data and the
effective field theory.

We observe a general trend of $K$ decreasing with increasing pore radius as
seen in Table~\ref{tabLL}. However,  the actual numerical values of the
Luttinger parameter $K$ for pores of varying radius can provide important
information on the sensitivity of the LL to perturbations coming from
commensuration effects or disorder; both of which are surely present in the
real experiments of Savard \emph{et al.}\cite{Savard:2009eq,Savard:2011fe}.
For a strictly one dimensional system, the introduction of a weak periodic
substrate that is commensurate with the density will only lead to complete
localization and the destruction of the harmonic fluid if $K >
1/2$\cite{Haldane:1981gd}.  Commensuration at other wavevectors is less
relevant and would require a greater value of $K$ to destabilize the LL.  The
introduction of a weak disorder potential, as might be present near the glassy
walls of the pore, is known to be relevant only when $K >
2/3$\cite{Giamarchi:1988dv}.

For the narrowest pores considered in this study $R< 3\ang$, we have
found a value of $K\approx 6$ at saturated vapor pressure, indicating
a strong tendency to form a solid, resulting from strong confinement and the
deep minimum and accompanying hardcore found in the effective interaction
potential $v_{\text{1d}}(z)$.  The experimental confirmation of this result
could be accomplished by noting that the formation of a quasi-solid should
impede superfluid flow through a helium-4 filled nanopore at low temperature.

It may be useful to compare the values of $K_{R}$ found here with other
studies of low dimensional helium.  For example, in a WA study of helium-4
confined to flow in the channels formed by screw dislocations with $R\sim3\ang$ in
solid helium, it was found that $K = 0.205(20)$\cite{Boninsegni:2007gc}, nearly
thirty times smaller than the comparable value of $K_{2.9}$ measured in this
study.  The sources of the discrepancy are rooted in the ``softness'' of the
confining potential inside the screw dislocation as helium atoms are able to
penetrate into the surrounding solid held at $\mu =0.02$~K corresponding to the
bulk melting point.

Much exciting work remains to be done in the nanopore system including a more
systematic study of the superfluid density which can be measured in bundles of
tubes or pores via torsional oscillator
techniques\cite{Wada:2009gp,Wada:2010jw}.  The construction of more realistic
models that contain both commensuration and disorder potentials would also
enhance the applicability of numerical simulations.  Further exploration of
the available parameter space is also in order, including altering the
chemical potential at fixed radius to simulate the effects of pressure which
can be freely tuned in experiments.  

In conclusion, we have studied a quantum fluid of bosonic helium-4 confined
inside nanopores of varying radii via large scale continuum Worm Algorithm
quantum Monte Carlo Simulations at saturated vapor pressure below the bulk
superfluid transition temperature.  The results show a progression of phases
inside the pore exhibiting a possible quasi-one-dimensional core surrounded by
concentric shells of helium depending on the radius.  When the core of the
nanopore has a non-zero density of helium, the finite temperature and scaling
properties of the density-density correlation function are fully described
within harmonic Luttinger liquid theory.  The description of the emergent low
energy phase of confined helium in terms of the harmonic field theory allows
for the extraction of the Luttinger parameter $K$ which is found to be a
decreasing function of radius.  As the pore radius increases, the inner
helium core is screened from the confining effects of the pore wall by the
surrounding matter, resulting in a pronounced enhancement of quantum
exchanges leading to superfluidity.   The precise relationship between the
material through which the pore has been sculpted and the resulting
Luttinger liquid parameters could be further explored leading to new
predictions and optimized experiments with the maximum likelihood of
detecting a universal and stable one-dimensional quantum harmonic fluid of
helium at low temperature.

\begin{acknowledgments}
The author would like to thank I.~Affleck, M.~Boninsegni and G.~Gervais for
many ongoing discussions.  This work was made possible through computational
resources provided by the National Resource Allocation Committee of Compute
Canada with all simulations taking place on Westgrid or SHARCNET.
\end{acknowledgments}

\end{document}